\documentclass[aps,onecolumn,amssymb]{revtex4} \usepackage{graphicx}

\def\be{\begin{equation}} \def\ee{\end{equation}}
\begin{document}
 
\title{On single-copy entanglement} \author{Ingo Peschel$^{1,*}$ and
Jize Zhao$^3$}

\affiliation{$^1$Max-Planck-Institut f\"ur Physik komplexer Systeme,
D-01187 Dresden, Germany\\
$^2$Institute of Theoretical Physics, Chinese Academy of Sciences,
Beijing 100080, China}

\vspace{0.5cm}

\begin{abstract}
 The largest eigenvalue of the reduced density matrix for quantum
 chains is shown to have a simple physical interpretation and
 power-law behaviour in critical systems.  This is verified
 numerically for XXZ spin chains.

\end{abstract}
\maketitle
 
\vspace{1cm}

In a recent paper Eisert and Cramer \cite{Eis05} investigated the
following problem from quantum information theory: Given an initial
state of a total system, what maximally entangled state between two
subsystems can be obtained from it by local transformations in both
parts, together with classical communication ?  The general answer is
that the spectrum of the reduced density matrix for the new state must
majorize that of the initial one \cite{Nielsen03}.  If the final state
is maximally entangled, it is the superposition of $M$ product states
which all have the same weight, and the reduced density matrix $\rho$
then has $M$ non-zero eigenvalues with magnitude $1/M$.  To reach this
state, the largest density-matrix eigenvalue $w_1$ in the initial
state must therefore be smaller than $1/M$.  This lead the authors to
study this eigenvalue and the associated quantity $S_1= -\ln w_1$ for
free-fermion chains.  For critical systems, they found a logarithmic
behaviour $S_1 \sim \ln L$ if the length $L$ of the smaller subsystem
diverges.  This is the same behaviour as for the entanglement entropy
$S= -{\rm tr}(\rho \ln \rho)$.  Moreover, the factor of
proportionality for the XX chain was 1/6 which is just half the value
one finds for $S$.  In the following we show that this can be
understood very easily and generalize the result to arbitrary
conformally invariant models.\\

Consider an infinitely long quantum chain in its ground state and an
interval of $L$ consecutive sites.  The reduced density matrix can be
written in the form \be \rho= \frac {1} {Z} \exp{(-H)}
   \label{eqn:rho1}
 \ee where $Z= {\rm tr}( \exp{(-H)})$ such that $\rho$ is normalized
 to one, ${\rm tr}(\rho) = 1$.  Then the largest eigenvalue of $\rho$
 is given by \be w_1= \frac {1} {Z} \exp{(-E_0)}
 \label{eqn:w1}
 \ee where $E_0$ denotes the smallest eigenvalue of $H$.  Therefore
 one has \be S_1= \ln Z + E_0
  \label{eqn:S11}
 \ee Now consider \be \ln{\rm tr}(\rho^n)=\ln{\rm tr}(\exp{(-n H)}) -n
 \ln Z \ee in the limit of large $n$.  Then only the smallest
 eigenvalue of $H$ contributes to the first term and one has \be
 \ln{\rm tr}(\rho^n)= -n E_0 -n \ln Z \ee Therefore the quantity $S_1$
 is given by the limit \be S_1= -\lim_{n \rightarrow \infty} {\frac
 {1}{n} \ln{{\rm tr}(\rho^n)}}
   \label{eqn:S13}
\ee By comparison, the entanglement entropy can be expressed as \be S=
\ln Z + < H >
 \label{eqn:Sa}
\ee and follows from values $n$ near one via \be S= - \frac {d} {dn}
{\rm tr}(\rho^n)|_{n=1}
  \label{eqn:Sb}
\ee Thus both $S$ and $S_1$ are determined by ${\rm tr}(\rho^n)$, or
equivalently ${\rm tr}(\exp{(-nH)})$.  The latter quantity has a
simple meaning in terms of the classical two-dimensional system
associated with the quantum chain \cite{PKL99,Calabrese04}.  For this
system, $\exp{(-H)}$ is the partition function for a plane with a cut
of length $L$ along which the variables are held fixed.  Thus ${\rm
tr}(\exp{(-nH)})$ gives the partition function of a system on a
$n$-sheeted Riemann plane which has the topology of a double winding
staircase.  Therefore $S_1$ is the difference between the
dimensionless free energies of one section of the Riemann surface and
of a normal plane.  The entropy $S$, on the other hand, is obtained
from an almost planar system of conical shape, and its logarithmic
behaviour can be traced back to the critical free energy of such
systems \cite{Cardy/Peschel88,Holzhey94,Calabrese04,Levine04}.\\

To determine $S_1$ for a critical system, one can now use the
conformal result \cite{Calabrese04} \be {\rm tr}(\rho^n)= b_n \,
(\frac {L} {a})^{\displaystyle{{-\frac {c} {6} (n-\frac {1} {n})}}}
\label{eqn:Spur}
\ee where $c$ is the central charge in the conformal classification
and $a$ denotes the radius of small circles around the branch points
which one has to exclude in the continuum model.  The constant $b_n$
is not determined from the conformal considerations but independent of
$L$.  Inserting (\ref{eqn:Spur}) into (\ref{eqn:S13}) one obtains
immediately \be S_1= \frac {c} {6} \ln (\,\frac {L} {a}\,) + k_1
   \label{eqn:S14}
\ee where a non-universal constant $k_1$ has been added.  The
logarithmic term is exactly one-half of the one appearing in $S$.  For
the XX chain, $c=1$ and one recovers the result found in \cite{Eis05}.
In the same way one can obtain $S_1$ for the other situations which
have been studied.  If the subsystem forms the end of a half-infinite
chain, the prefactor of the logarithm is reduced by 2 and if one
considers a chain of length $L$ which is divided into two sections of
length $l$ and $L-l$, the result is \cite{Calabrese04} \be S_1= \frac
{c} {12} \ln [\,\frac {2 L} {\pi a} \sin(\frac {\pi l} {L})\,] + k_1
  \label{eqn:S15}
\ee Although the conformal results give the general answer to the
problem, it is still instructive to look at the free-fermion case once
again.  Then the operator $H$ in (\ref{eqn:rho1}) is also fermionic
and has the diagonal form \be H= \sum_{k} \varepsilon_k
c_k^{\dagger}c_k
   \label{eqn:H}
\ee where the single-particle eigenvalues $\varepsilon_k$ follow from
the correlation functions in the considered state
\cite{Peschel03,Vidal03}.  For a half-filled chain of fermions,
corresponding to an XX model, these quantities have been studied
before \cite{Cheong/Henley03,Peschel04} and appear in pairs
$(\varepsilon,-\varepsilon)$ if $L$ is even.  Then one can express $S$
in the form \be S= 2\,\sum_{\varepsilon_k >0} \left [ \,
\ln{[1+\exp{(-\varepsilon_k)}]}+ \frac {\varepsilon_k}
{\exp{(\varepsilon_k)}+1} \right ]
   \label{eqn:Sc}
\ee whereas $S_1$ is given only by the first sum.  For odd $L$, an
additional eigenvalue $\varepsilon_k= 0$ appears and gives a
contribution $\ln 2$ to both $S$ and $S_1$.  This is valid for
arbitrary $L$.  In the limit $L \rightarrow \infty$, however, the
situation simplifies as the spectrum of the $\varepsilon_k$ then
becomes linear and dense.  For even $L$ it is given by \be
\varepsilon_k= \frac {\pi^2} {2 \ln L} (2k+1); \;\; k=0,1,2,...
  \label{eqn:eps}
\ee One can then go over to integrals in (\ref{eqn:Sc}) and finds that
the two contributions to $S$ are exactly the same (and proportional to
$\ln L$).  This gives directly $S_1 = S/2$ for the leading logarithmic
terms.  The argument can be extended to non half-filled systems, where
the spectrum of the $\varepsilon_k$ is shifted.\\

In the free-fermion case one can also make use of the results for the
$\varepsilon_k$ obtained for non-critical XY spin chains and for the
Ising model in a transverse field \cite{PKL99,Peschel04/2,Its05} to
calculate $S_1$.  The same holds for the XXZ model for $\Delta > 1$.
As the entanglement entropy, $S_1$ is then finite and diverges only as
one approaches the critical point.  Using the formulae in
\cite{Peschel04/2}, it can be expressed in closed form.  For example,
a transverse Ising chain divided into two halves gives in the
disordered region \be S_1= \frac {1} {24} \left[ \ln (\frac {16} {k^2
k'^2}) -\pi \frac {I(k')} {I(k)} \right ]
 \label{eqn:elli}
\ee where $k$ measures the coupling, $k'^2=1-k^2$ and $I(k)$ is the
complete elliptic integral of the first kind.  Near criticality ($k
\rightarrow 1$) this has the expansion \be S_1= \frac {1} {24} \left [
\ln (\frac {8} {1-k}) -\frac {\pi^2} {2} \frac {1} {\displaystyle{\ln
(\frac {8} {1-k})}} \right ]
 \label{eqn:crit}
\ee The logarithms can be expressed in terms of the correlation
length, since $\xi \sim 1/(1-k)$.  This expansion differs from that
for $S$, where the second term does not appear.  For the XY chain one
obtains similar results.\\

Returning to $w_1$ itself, the results imply that it varies as a power
in $1/L$ for critical systems.  Its vanishing for $L \rightarrow
\infty$ is connected with the fact that the spectrum of $\rho$, as
well-known from DMRG calculations, flattens more and more as $L$
becomes larger.  The power-law behaviour can also be seen in numerical
calculations.  This is shown in Fig.1 where results for XXZ spin
chains with anisotropies $\Delta$ between $-1$ and $+1$ are presented.
They were obtained from DMRG calculations on open chains of odd total
length up to 801 sites, keeping m=500 states but using no sweeps.  The
division was into subsystems of $L$ and $L-1$ sites with $L$ odd.
Thus one is dealing with the situation described by the formula
(\ref{eqn:S15}).  The odd length was chosen because then the
convergence is better.  As one can see, the curves in the
double-logarithmic plot become rather straight for large $L$ and
differ by additive constants.  The resulting values of $c$ are shown
in the right part of the figure as function of $1/\ln L$.  They
clearly tend to the value 1, but the convergence is relatively slow
and dependent on the value of $\Delta$, as well as on the chosen
expression for the slope of $S_1$.  A second order extrapolation in
$1/\ln L$ gives $c$-values which deviate from $1$ by at most $5\%$.
By contrast, the values for $c$ determined from the entanglement
entropy $S$ lie much closer to 1 and converge faster.  One could
associate this with the different behaviour of the two quantities near
criticality mentioned above.\\
\begin{figure}
\includegraphics[width=8cm]{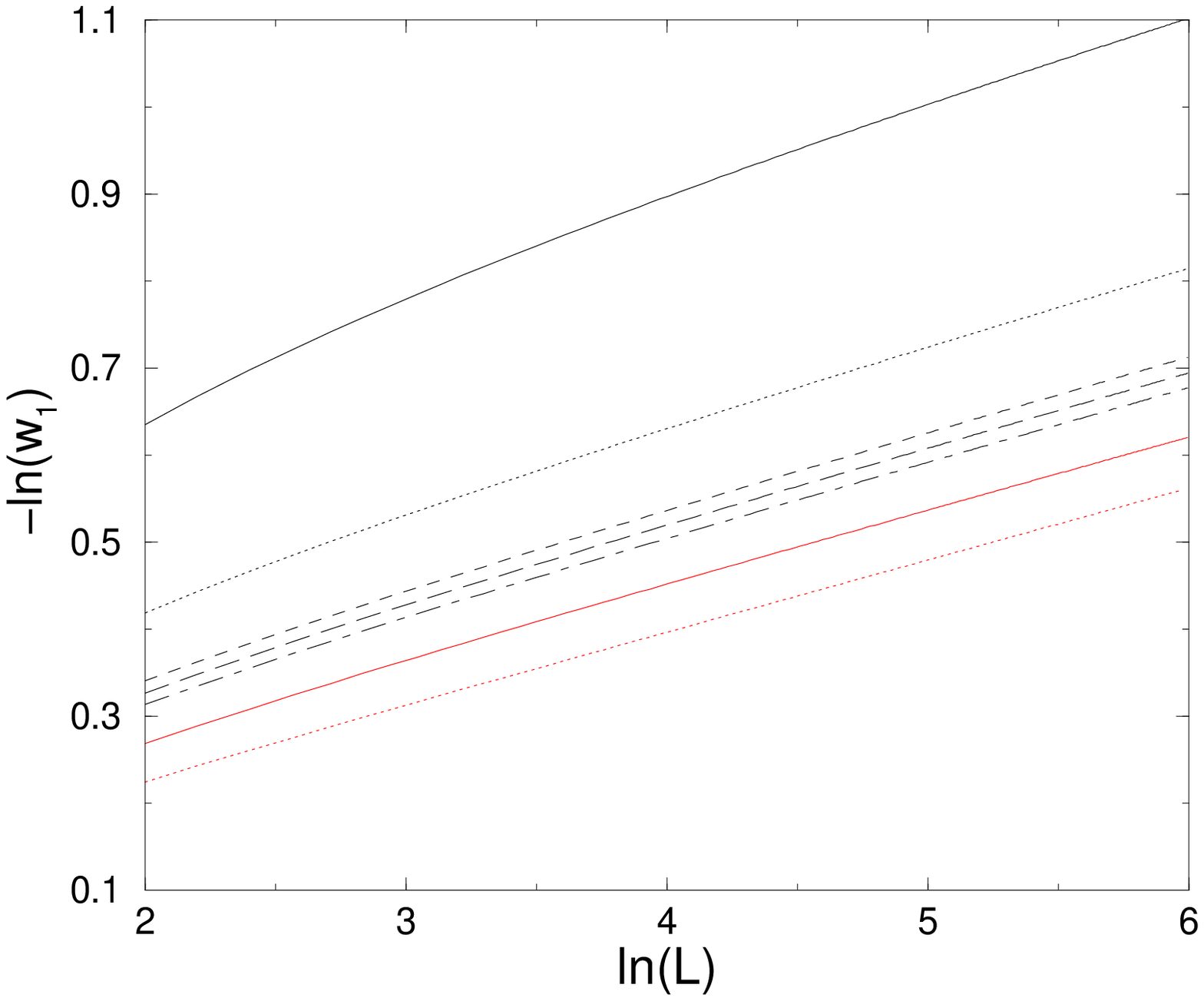}
\includegraphics[width=8cm]{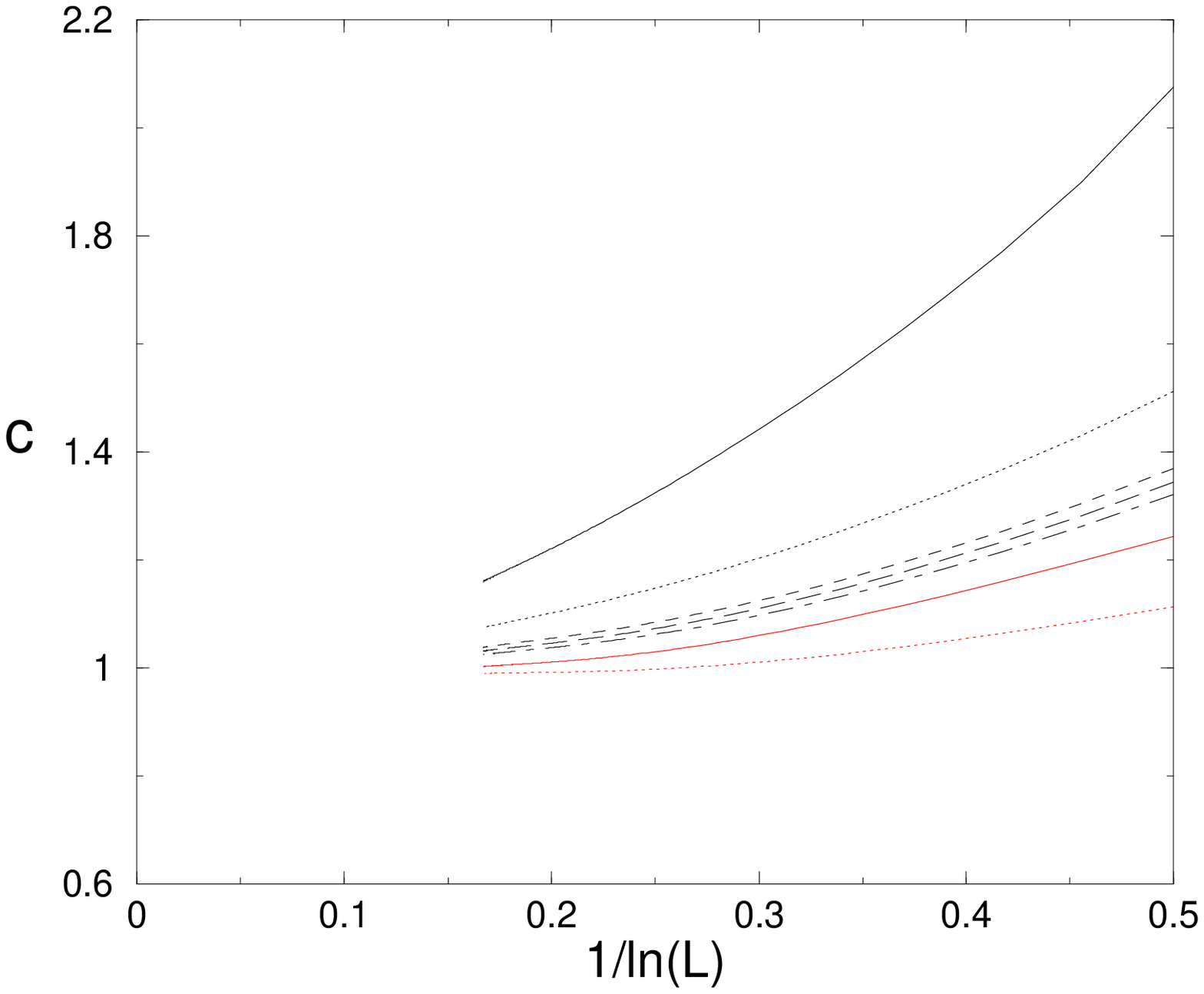}
\vspace{1mm} \caption{Left: Largest eigenvalue of $\rho$ for XXZ
chains with half-length L. Right: Central charge $c$ determined from
the slopes in the left figure multiplied by 12.  The curves correspond
to the anisotropies $\Delta=-0.9;-0.5;-0.1;0;0.1;0.5;0.9$, from top to
bottom.}
\label{fig1.eps}
\end{figure}
Summing up, we have shown that the largest density-matrix eigenvalue
for a quantum chain can be interpreted as a difference of two free
energies.  Correspondingly, it shows universal features if the system
is critical.  For the problem cited at the beginning, this means that
the rank $M$ of the maximally entangled state grows unlimited with the
subsystem size (although the power is relatively small) and the
central charge $c$ appears again in the field of quantum
information.\\
 
{\it Note added:} The conformal result for $S_1$ was given
independently also by Or\'{u}s et al.\cite{Orus05} with an additional
subleading term.

$^*$ Permanent address: Fachbereich Physik, Freie Universit\"at
Berlin, Arnimallee 14, D-14195 Berlin, Germany

\end{document}